\documentclass{article}
\usepackage{threeparttable}
\usepackage{spconf,amsmath,graphicx}
\usepackage{float}
\usepackage{subfigure}
\usepackage{graphicx}
\usepackage{multirow}
\usepackage{booktabs}
\usepackage{tabularx}
\usepackage{hyperref}

\title{S-DCCRN: Super Wide Band DCCRN with learnable complex feature for speech enhancement}
\name{Shubo Lv$^{1}$%\thanks{Thanks to XYZ agency for funding.}
, Yihui Fu$^{1}$, Mengtao Xing$^{1}$, Jiayao Sun$^{1}$, Lei Xie$^{1}$, Jun Huang$^{2}$, Yannan Wang$^{2}$, Tao Yu$^{2}$}
\address{Audio, Speech and Language Processing Group (ASLP@NPU), \\ Northwestern Polytechnical University, Xi'an, China\\
Tencent Ethereal Audio Lab, Tencent Corporation, Shenzhen, China}
\begin{document}
\ninept
\maketitle
\begin{abstract}
  In speech enhancement, complex neural network has shown promising performance due to their effectiveness in processing complex-valued spectrum. Most of the recent speech enhancement approaches mainly focus on wide-band signal with a sampling rate of 16K Hz. However, research on super wide band (e.g., 32K Hz) or even full-band (48K) denoising is still lacked due to the difficulty of modeling more frequency bands and particularly high frequency components.
  %As the sampling rate increases, the immersion of the sound quality will rise. Meanwhile, the difficulty of modeling different frequency bands will also increase.
  In this paper, we extend our previous deep complex convolution recurrent neural network (DCCRN) substantially to a super wide band version -- \textit{S}-DCCRN, to perform speech denoising on speech of 32K Hz sampling rate.
  We first employ a cascaded sub-band and full-band processing module, which consists of two small-footprint DCCRNs -- one operates on sub-band signal and one operates on full-band signal, aiming at benefiting from both local and global frequency information. Moreover, instead of simply adopting the STFT feature as input, we use a complex feature encoder trained in an end-to-end manner to refine the information of different frequency bands. We also use a complex feature decoder to revert the feature to time-frequency domain. Finally, a learnable spectrum compression method is adopted to adjust the energy of different frequency bands, which is beneficial for neural network learning. The proposed model, \textit{S}-DCCRN, has surpassed PercepNet as well as several competitive models and achieves state-of-the-art performance in terms of speech quality and intelligibility. Ablation studies also demonstrate the effectiveness of different contributions.
\end{abstract}
\begin{keywords}
speech enhancement, super wide band, \textit{S}-DCCRN
\end{keywords}
\vspace{-0.5cm}
\section{Introduction}
\label{sec:intro}
\vspace{-0.2cm}

% Speech enhancement typically focuses on extracting a high-quality version of a target speaker’s utterance from the mixture that contains the target speaker in addition to multiple competing ambient sounds. After processing, severe noise and reverberation can be alleviated.

% Thanks to the success of deep learning in recent years, speech enhancement has received remarkable progress. For a long time, the researches on speech enhancement mainly focus on wide-band scenarios (sampling rate = 16000 Hz). For frequency domain, convolution recurrent network (CRN)~\cite{tan2018convolutional} is one of the most popular convolution encoder-decoder (CED) structures for speech enhancement. For time domain, Conv-TasNet is a powerful model for speech enhancement that using a combination of trainable analysis/synthesis filter-banks and a mask prediction network.
%添加一些16k的典型文章引用 简要讲一下方法
With the fast development of tele-conference and other real-time speech communication scenarios, the demand for high-quality Hi-Fi speech has increased sharply. With a higher sampling rate, speech will contain richer information and more fine details, especially in higher frequency bands. However, most current deep learning based speech enhancement approaches mainly focus on wide band signal at sampling rate of 16K Hz. The potential of speech enhancement on super wide band~\cite{cox2009itu} or even full-band signal is still to be explored since the challenges exist in modeling more frequency bands and particularly high frequency components. Moreover, modeling with larger dimensional features will cause higher complexity of the modeling, making real-time implementation becomes more difficult. Some speech enhancers have adopted compressed features like bark spectrum~\cite{valin2018hybrid} to model high frequency signal while feature compression may unavoidably lose important information of frequency bands, resulting in sub-optimal performance.

%challenge是啥呢？读者就是想看你的challenge呢？
%不是说多band就challenge了。。。输入多我模型大不就行了？你好好想想你坐这么久 难点在哪？至少要三四句介绍挑战

% the high quality of the speech in our audio calls is a need during these times as we try to stay connected and collaborate with people every day. The quality and immersion of super wide-band speech are better than wide-band. At the same time, the challenge of this scenario is harder. A key challenge in developing a speech enhancer for super wideband speech is how to perceive the different frequency bands of speech, especially high frequency, while greatly suppressing the noise interference. Furthermore, the increase of frequency bands requires a large number of neurons and weights, resulting in high complexity.

% The recent deep noise suppression challenge (DNS) series\cite{ reddy2021interspeech} have one track on fullband (sampling rate = 48000 Hz) scenarios, which focused on real-time denoising.

% Recent advances in deep learning have enabled significant progress on speech enhancement problems. Typical methods tackle the problem by learning a spectral mapping or masking on the magnitude spectrum, while the inverse STFT process to recover waveform introduces audible artifacts due to missing or mismatching phase.

For a long time, DNN-based speech front-end algorithms attempt to only enhance the noisy magnitude while the noisy phase is directly incorporated for speech waveform reconstruction.
The reasons can be attributed to the unclear structure of the phase, which is considered challenging to estimate.
Subsequently, complex ratio mask (CRM)~\cite{williamson2015complex} was proposed by Williamson \textit{et al.}, which can reconstruct speech perfectly by enhancing both real and imaginary components of the noisy speech simultaneously.
Wide-band scenario based SOTA methods like SDD-Net~\cite{li2021simultaneous} and DCCRN~\cite{hu2020dccrn} have shown outstanding performance, especially for complex denoising cases with low SNRs and sudden noise. DCCRN combines the advantages of both DCUNET~\cite{choi2018phase} and CRN~\cite{tan2018convolutional}, using LSTM to model temporal context with significantly reduced trainable parameters and computational cost. SDD-Net applies the power-compressed spectrum~\cite{li2021importance} as the input features and employs four particularly designed stages, which dramatically improved speech quality in simultaneous dereverberation and denoising.
%咋又不说具体咋做的？为啥效果就好呢？为啥就sota了呢？直接抄原文contribution就行了
For full-band scenario, studies in RNNoise~\cite{valin2018hybrid} adopt bark spectrum, instead of STFT, as the input of the model. Bark spectrum only has 22 dimensions on frequency axis totally~\cite{moore2012introduction}, which can reduce the model size to a great extend and speed up model inference.
%divide the spectrum into the same approximation of the Bark scale. This results in a total of 22 bands instead of full frequency bands which can smaller model size and speed-up inference.
The bark spectrum assumes that the spectral envelopes of the speech and noise are flat sufficiently~\cite{valin2018hybrid}. %（这句话位置放错了 应该放上面刚介绍bark时候就说）
However, this method may lead to severe attenuation in the real acoustic scenario due to the complexity of the real acoustic scenario (such as sudden noises and reverberation), which leads to excessive noise residual. %In addition, the original frequency band will be compressed through this conversion, and the phase information is discarded. %(假设怎么能只因为噪声不同就有缺陷呢？难道不应该是因为环境复杂导致效果不理想吗)
%Proposed by xxx, RNNoise relies on proven signal processing techniques and uses deep learning to replace the estimators that have traditionally been hard to tune correctly. (什么signal processing，estimator是啥。得说明白这东西具体咋做的 好处是啥 劣势又是啥（可说可不说）。 我没看过rnnoise我完全不知道这都是啥模块 千万不要给人一种听君一席话浪费三秒钟的感觉 )
Very recently, the PercepNet~\cite{valin2020perceptually} proposed a perceptual band representation which operates on only 32 triangular spectral bands, spaced according to the equival entrectangular bandwidth (ERB) scale~\cite{moore2012introduction}. %（同上 难道这个模型创新点仅仅是帧长帧移？）
However, the resolution of bark scale and ERB scale are more rough than linear spectrum from STFT, leading to the leakage of information of frequency bands. Speech enhancement on super wide band/full-band signal has drawn much attention recently -- deep noise suppression challenge (DNS)~\cite{reddy2021interspeech} has particularly set up a full-band track.

% Although complexity being much lower than the maximumal-lowed by the recently concluded first DNS challenge, PercepNet ranked second in the real-time track.

This paper proposes \emph{super wide band} DCCRN (\textit{S}-DCCRN) for speech enhancement in super wide-band scenarios at 32K Hz sampling rate. The contribution of this work is three-fold, evaluated objectively on Voicebank and Demond dataset and subjectively on DNS-2021 blind test set.

%(介绍自己工作全用现在时)
We propose two lightweight DCCRN sub-modules for \emph{sub-band and full-band} (SAF) modeling respectively, since it is considered that low frequency bands contain higher energy while higher frequency bands have a great impact on subjective perception~\cite{takahashi2017multi}.
Therefore, sub-band processing module is employed to model low frequency bands and high frequency bands separately. However, only adopting sub-band processing may cause unsmooth connection among frequency bands since there is no explicit information interaction between low- and high-frequency components. Thus, we further apply a full-band processing module to smooth the boundaries of different frequency bands.
%which can predict fullband features of a clean target from those of noisy input. （这句话又感觉说了又没说 你得要说子代有啥特性，所以为啥要划子代）
In detail, the sub-band processing module consists of a sub-band DCCRN, which substitute the complex convolution of original DCCRN with group complex convolution to model low-frequency bands and high-frequency bands separately.
% The difference between fullband processing and subband processing is that we replace group complex convolution with complex convolution.
% 为啥要respectively？是不是得说两个包含的信息不一样？分别建模有利于模型关注不同频带的特性？
The convolution pathways~\cite{lv2021dccrn+} among encoder and decoder of both full-band and sub-band processing modules are used for better information interaction avoiding information lost in full-band.
%pathway是啥？全都是第一次出现的词汇，得说明一下动机作用
%（为啥？to xxx）
With a smaller model size, the SAF module leads to 0.17 PESQ improvement compared with the oracle DCCRN.
%（which指代不明 到底是说现在模型还是dccrn？）

Inspired by spectrum compression in wide-band denoising~\cite{li2021importance}, we introduce \emph{learnable spectrum compression} (LSC) in our model, which can dynamically adjust the energy of different frequency bands. The use of LSC results in more clear patterns on the high-frequency bands and this update brings an extra PESQ gain of 0.07.

Motivated by the encoder/decoder block of the DPT-FSNet~\cite{dang2021dpt}, we employ a \emph{complex feature encoder} (CFE) after STFT and a \emph{complex feature decoder} (CFD) before iSTFT. We keep the same STFT points as most wide-band speech enhancement models. Although the frequency resolution is relatively low for high sampling rate scenario, the CFE block can refine the information of different frequency bands of the STFT spectrum. With learnable spectrum compression, this update brings an extra PESQ gain of 0.07.

The proposed \textit{S}-DCCRN model surpasses all tested %several（那就是还有比你强的？那搞这个模型干啥）
SOTA models, including RNNoise and DCCRN, and obtains superior performance with 3.62 MOS score on the blind test set of Interspeech 2021 DNS challenge~\cite{reddy2021interspeech}.
\vspace{-0.3cm}

% This paper is organized as follows. Our proposed S-DCCRN is introduced in Section 2. In section 3, the experiments and the results are presented. Lastly, in section 4 we conclude out work.

\section{Super Wide Band DCCRN (S-DCCRN)}
\label{sec:format}
    %%%%%%%%%%%%%总图
    \begin{figure}[t]
    \centering
    \vspace{-0.2cm}
    \includegraphics[width=1.0\linewidth]{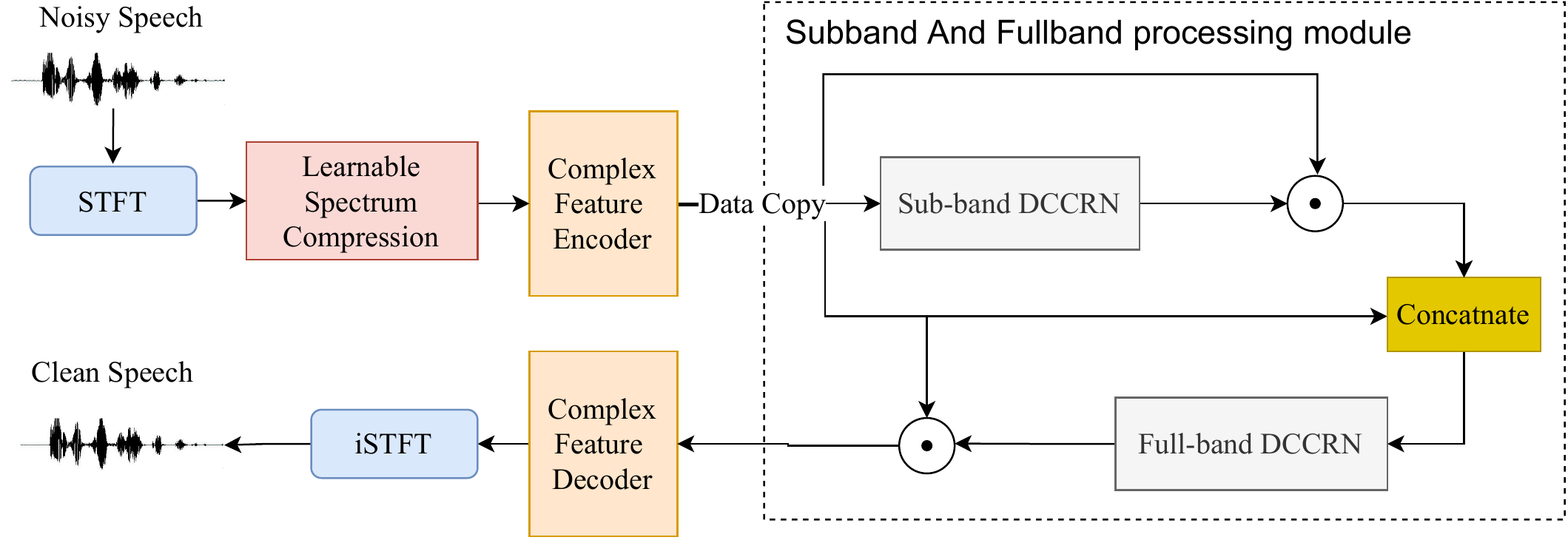}
    \vspace{-0.5cm}
    \caption{Network structure of the proposed \textit{S}-DCCRN}
    \label{fig:net}
    \vspace{-0.7cm}
    \end{figure}
    %%%%%%%%%%%%%%%%%%%%%%%
%Figure~\ref{fig:net} gives the block diagram of the neural network model used for Net. To model different frequency bands, especially high frequency, we combine sub-band and full-band processing. Furthermore, we design a complex feature encoder and decoder network that can extract the useful feature map of the STFT feature. Finally, Considering the difference in energy of bands, we propose a learnable spectrum compressing method to enhance the network's perception of different frequency bands.
\vspace{-0.3cm}
\subsection{Complex Feature Encoder/Decoder}
\vspace{-0.1cm}
Researchers usually adopt Bark spectrum as the input of the network for full-band speech enhancement in wide band scenario, which can convert physical frequency to psychoacoustic frequency based on human perception~\cite{valin2018hybrid}. However, the original physical frequency bands are compressed through this conversion, and the phase information is discarded as well. In addition, features based on the human perception may not be suitable for the input of network. %啊？为啥？模型增强出来不就是给人听的吗？
On the other hand, using the STFT features directly also causes certain problems.
%为啥是however 哪来的转折关系？
With a larger number of points of STFT, the network complexity will increase due to high-dimensional input features that are hard to model. On the contrary,  %为啥长了就增加复杂度了？是不是因为维度高了输入复杂了难以建模了？
the use of the STFT features with a smaller number of points can also cause the degradation of frequency resolution.
%多用名词和被动式 少要用主动的句子表示。这块给你改了
In this paper, inspired by the encoder/decoder block of DPT-FSNet~\cite{dang2021dpt}, we adopt a complex feature encoder/decoder after STFT to refine the information of different frequency bands based on 512-dimensional complex STFT features.
%你用中文说一下想说什么意思？（答：就是用这个cfe/cfd来细化STFT复数谱中不同频带的信息）
%%这句没改吗？
% based 512-dimensional complex STFT features by learning. 这啥意思

%为什么全是will 怎么怎么 都得是现在是
    % %%%%%%%%%%%%%总图
    % \begin{figure}[t]
    % \centering
    % \includegraphics[width=1.0\linewidth]{picture/FDNet.pdf}
    % %\vspace{0.5cm}
    % \caption{Illustration of the Network structure of the proposed S-DCCRN, (b) Complex Feature Encoder, (c) Complex Feature Decoder and (d) Group ComplexConv}
    % \label{fig:net}
    % %\vspace{-0.6cm}
    % \end{figure}
    % %%%%%%%%%%%%%%%%%%%%%%%
    %%%%%%%%%%%%%CFE_CFD
    \begin{figure}[t]
    \centering
    \includegraphics[width=.9\linewidth]{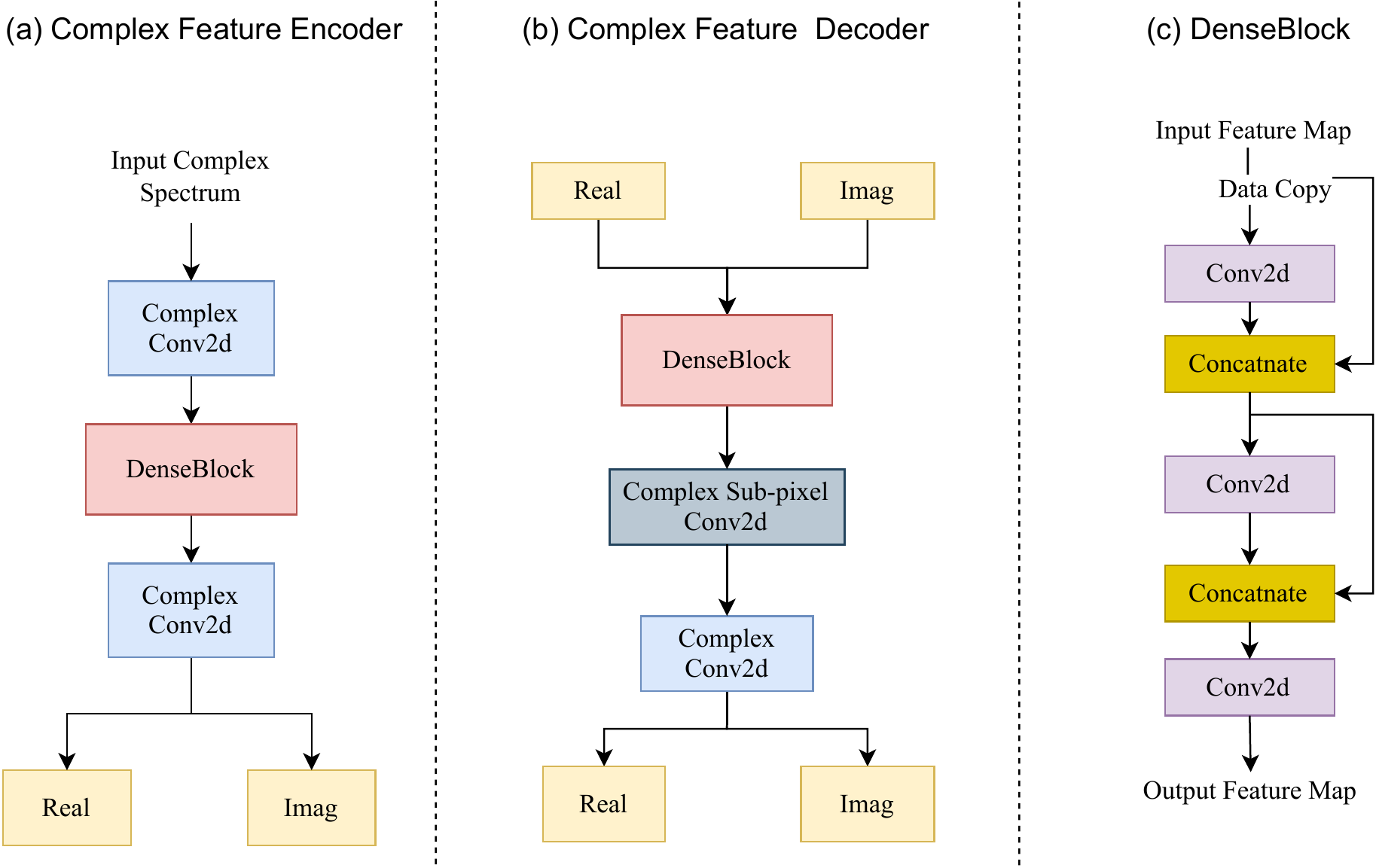}
    %\vspace{0.5cm}
    \vspace{-0.2cm}
    \caption{Complex feature encoder/decoder (CFE/CFD) module}
    \label{fig:cfed}
    %\vspace{-0.7cm}
    \end{figure}
    %%%%%%%%%%%%%%%%%%%%%%%
As shown in Figure~\ref{fig:cfed} (a), the input of the complex feature encoder (CFE) module is the T-F spectrum
%resulting from short-time Fourier transform (STFT).
obtained by STFT.
We employ complex conv2d with a kernel size of 1 to extract high-dimensional information. Then a dilated dense block~\cite{pandey2020densely} is used to capture long-term contextual features from time scale. Finally, a complex conv2d is adopted to extract complex local features. LayerNorm %~\cite{ba2016layer}
and PReLU activation %~\cite{he2015delving}
are successively performed after each convolution.

As shown in Figure~\ref{fig:cfed} (b), the input of the complex feature decoder (CFD) module is the real/imag features of the output of SAF module. In detail,
%the complex feature decoder (CFD) module is similar to CFE module.
%same就完了？那这俩不是一个东西吗？有啥不同？
%咋没改呢？光similar不就一模一样了？我看fig2 a和b也不一样呀？
we employ the dilated dense block to process the estimated real/imag features. Then the output of dilated dense block is processed by complex pixel convolution, which substitute the convolution in the pixel convolution~\cite{shi2016real} with complex convolution. The pixel convolution is considered as a better alternative for transposed convolution to avoid checkerboard artifacts~\cite{odena2016deconvolution}. Finally, a complex convolution is performed to revert the high-dimensional feature to the time-frequency domain.
%这块应该说一下pixel conv的优势 为啥不用反卷积之类的？
As shown in Figure~\ref{fig:cfed} (c), each dense block consists of five layers of conv2d. The convolutions across the frames are causal. The dense connection to all the previous layers avoids the vanishing gradient problem~\cite{pandey2020densely}.
\vspace{-0.3cm}

\subsection{Sub-band and Full-band Processing Module}
\vspace{-0.2cm}
As the sampling rate increases, the number of frequency bands also increases to a great extent. Different frequency bands are hard to be modeled by only full-band processing because the information among low-frequency and high-frequency is substantially different~\cite{takahashi2017multi}. It is not optimal to model them within one module. On the other hand, sub-band processing can cause a certain unsmooth connection at the boundary of different frequency bands since there is no information interaction among different bands. Based on the considerations mentioned above, we propose a sub-band And full-band processing (SAF) module to take the advantage from both. % to combine them.

As shown in Figure~\ref{fig:net}, we use the encoded feature from CFE as the input of the SAF module, which is mainly composed of a sub-band DCCRN and a full-band DCCRN. In the SAF module, the features are firstly processed by a sub-band DCCRN. The concatenation of the sub-band DCCRN output together with the encoded feature from CFE, which is considered to help to smooth frequency bands, is treated as the input of the full-band DCCRN. The output of full-band DCCRN is the complex ratio mask (CRM) of the encoded feature from CFE.

    %%%%%%%%%%%%%总图
    \begin{figure}[t]
    \centering
    \includegraphics[width=.7\linewidth]{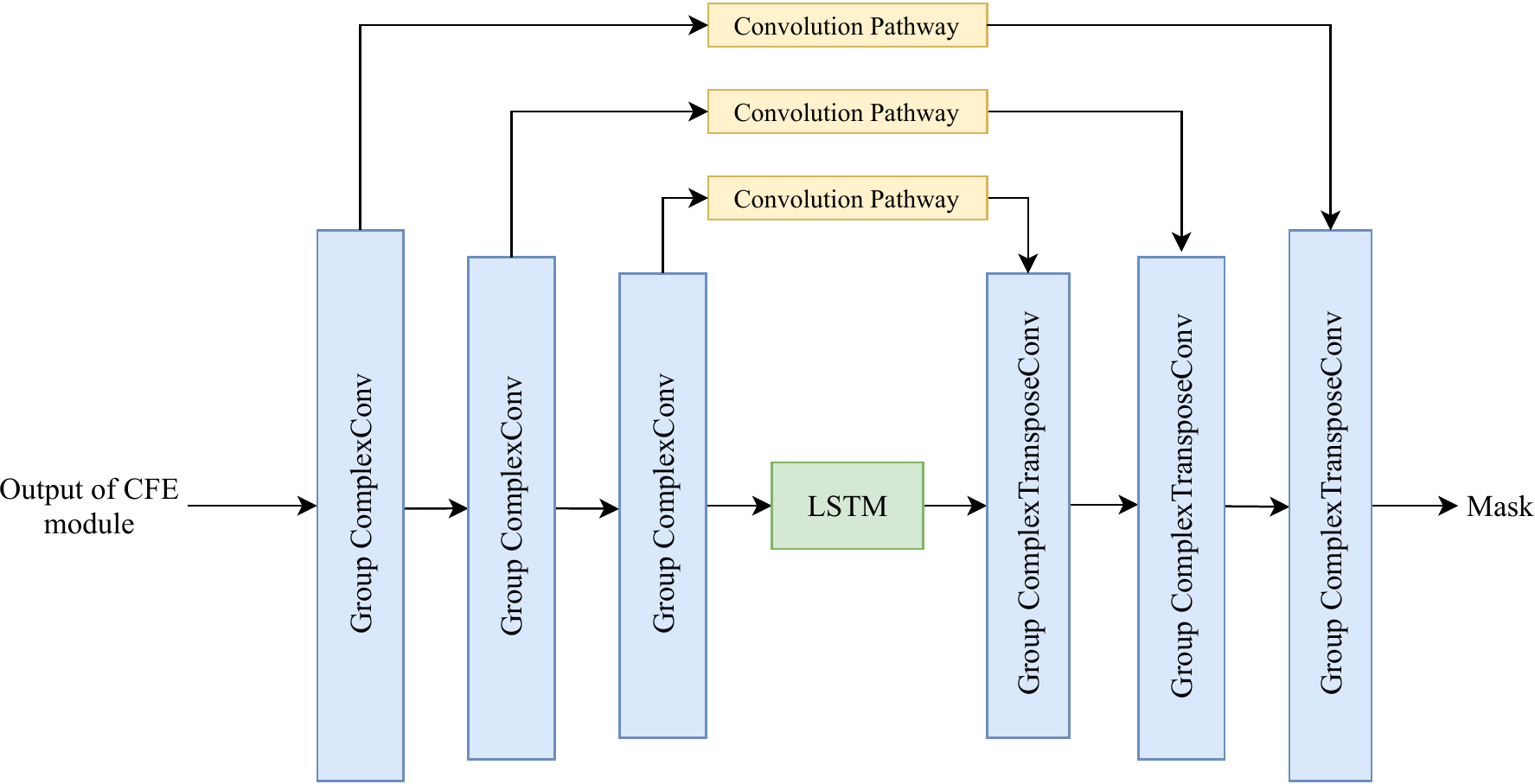}
    \vspace{-0.4cm}
    \caption{The sub-band DCCRN module}
    \label{fig:subdccrn}
    \vspace{-0.6cm}
    %\vspace{-0.5cm}
    \end{figure}
\begin{figure}[t]
    \centering
    \vspace{-0.2cm}
    \includegraphics[width=.5\linewidth]{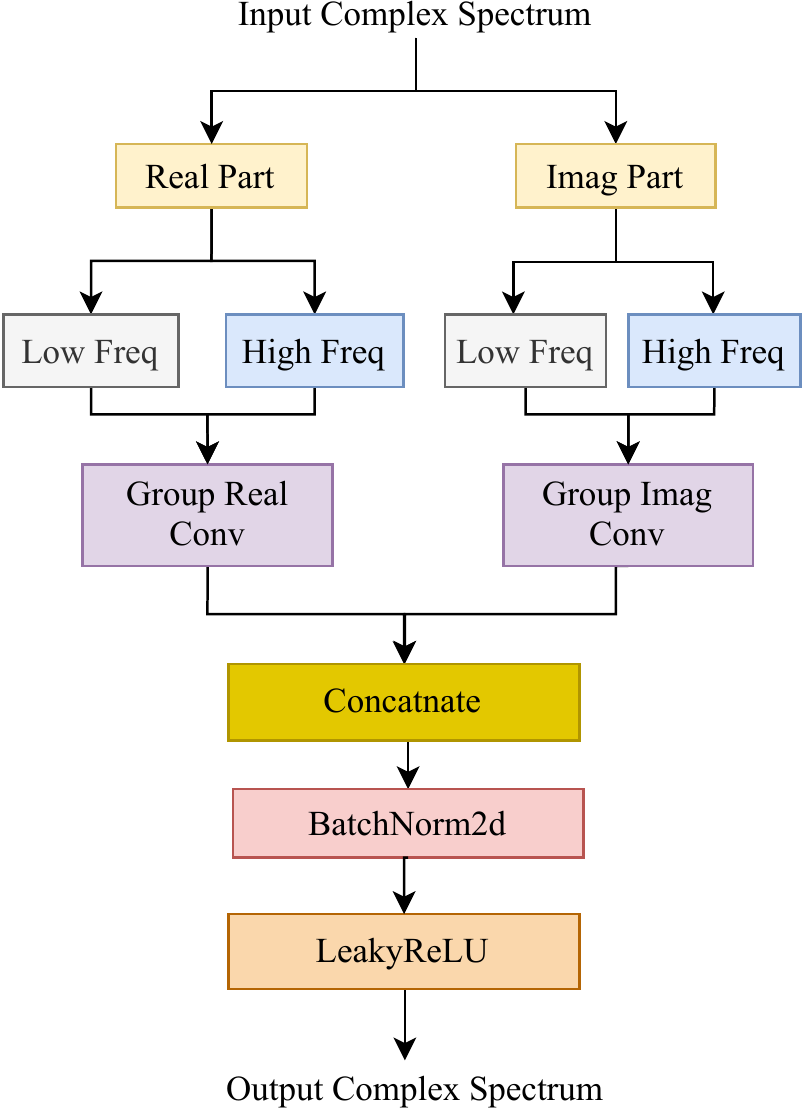}
    \vspace{-0.2cm}
    \caption{Complex group convolution}
    \label{fig:groupconv}
    \vspace{-0.6cm}
    \end{figure}

    %%%%%%%%%%%%%%%%%%%%%%%
The structure of the sub-band DCCRN is shown in Figure~\ref{fig:subdccrn}. The general design of the sub-band DCCRN is similar to the oracle DCCRN, but the complex convolution block in oracle DCCRN is substituted with a complex group convolution block. As shown in Figure~\ref{fig:groupconv}, complex group convolution block aims to model low-frequency bands and high-frequency bands separately. In addition, to aggregate richer information from each encoder layer and relieve the unsmooth connection among frequency bands, we employ the convolution pathway block between encoder and decoder, which was proven useful in DCCRN+~\cite{lv2021dccrn+}. Specifically, the convolution pathway between encoder and decoder consists of a complex convolution block and batch normalization.
%这咋没介绍pathway的结构呢？

After concatenating the encoded feature from CFE together with the output of sub-band DCCRN, we use another full-band DCCRN, which takes ordinary complex convolution instead of complex group convolution, to further regenerate and smooth different frequency bands. Full-band DCCRN also employs the convolution pathway among encoder and decoder for better information interaction.
    %%%
    %%%
\vspace{-0.4cm}
\subsection{Learnable Spectrum Compression}
\vspace{-0.1cm}
It is pointed out that local patterns in the spectrum are often different in each frequency band: the lower frequency band tends to contain high energies, tonalities as well as long sustained sounds, while the higher frequency band is likely to have low energy components, noise, and rapidly decaying sounds~\cite{takahashi2017multi}. Recently, spectrum compression on wide-band denoising has shown promising results, which can increase the energy of the high-frequency bands through a compression rate of 0.5~\cite{li2021importance}.

We believe that the compression rate of frequency bands should be different since high-frequency bands may require a lower compression ratio to maintain its high energy. This inspires us to develop a learnable spectrum compression module using a set of network layers to compress the STFT spectrum. The sigmoid activation is performed after the learnable network layers, aiming to compress the output to 0 $\sim$ 1.
In detail, the learnable spectrum compression can be described as
\vspace{-0.1cm}
{
    \setlength\abovedisplayskip{1pt}
    \setlength\belowdisplayskip{0.5pt}
    \begin{equation}\label{lsc}
    \vspace{-0.2cm}
    \begin{aligned}
    Y_{\textbf{LSC}} =  \left | Y \right | ^ \alpha e^{j\varphi _{Y}}
    \end{aligned}
    \end{equation}
}where ${Y}$ and ${\alpha}$ denote the noisy spectrum and the learnable parameters respectively.
%neural layers
%It should be noted that the dimensions of spectrum compression layers is 256.
\vspace{-0.3cm}
\subsection{Loss Function}
For the learning objective, we first apply SI-SNR~\cite{luo2019conv} loss, which is a time-domain loss function.
% that could be defined by the following formula:
% \begin{equation}
% % \left \{
%      \vspace{-0.2cm}
%      \setlength{\arraycolsep}{0.3pt}
%      \begin{cases}
%     \boldsymbol{s}_{\text{target}}&=(<\Tilde{\boldsymbol{s}},\boldsymbol{s}>\cdot \boldsymbol{s})/||\boldsymbol{s}||_{2}^{2} \\
%      \boldsymbol{e}_{\text{noise}}&=\boldsymbol{\Tilde{s}}-\boldsymbol{s_{\text{target}}}\\
%      \mathcal{L}_{\textbf{SI-SNR}} &=10\log10(\dfrac{||\boldsymbol{s_{\text{target}}}||_{2}^{2}}{||\boldsymbol{e_{\text{noise}}}||_{2}^{2}}) \\
%      \end{cases}
%       %\vspace{-0.2cm}
% % \right.
% \end{equation}
%说一下每个符号的含义？
%咋没改呢？
Furthermore, we employ complex mean-squared error (MSE) loss and the Kullback-Leibler Divergence~\cite{polani2013kullback} to improve the similarity between the estimated spectrum and the clean spectrum in complex domain. The purpose of KL Divergence is to optimize clean and estimated spectrum from the perspective of probability distribution. The three losses are optimized jointly by:
%而且没说明为啥要用这三个loss 意义何在？improve the similarity between the estimated spectrum and the clean spectrum 这所有loss不都是这个功效吗？
\vspace{-0.3cm}
\begin{equation}
    \vspace{-0.4cm}
% \left \{
     \setlength{\arraycolsep}{0.3pt}
     \begin{cases}
     \mathcal{L}_{\textbf{cMSE}} &= \frac{1}{T\times F}\sum_{t,f}^{}\left |      \left | X \right | e^{j\varphi _{X}} - \hat{\left |  X  \right |} e^{j\varphi _{\hat{X}}}\right | \\
     \mathcal{L}_{\textbf{KL}} &=\frac{1}{T\times F} \sum_{t,f}^{}\hat{X}\cdot log(\frac{\hat{X}}{X})\\
     \mathcal{L} &= \mathcal{L}_{\textbf{SI-SNR}} + \mathcal{L}_{\textbf{cMSE}} + \mathcal{L}_{\textbf{KL}} \\
     \end{cases}
% \right.
\end{equation}
%注意公式后面也要有标点符号。因为公式也是句子的一部分
%sisnr+cmse+dkl。。。咋能用几个单词来表示变量呢？看我咋写的
where ${\hat{X}}$ and ${X}$ denote the network output and clean spectrum respectively. We omit the dependency of the target speech spectral bins${X_{t,f}}$ on the frequency and time indices ${t,f}$ for brevity.
\vspace{-0.3cm}
\section{Experiments}
\vspace{-0.3cm}
\label{sec:print}
\subsection{Datasets}
\vspace{-0.2cm}
We carry out speech enhancement experiments on audio samples with 32K sampling rate. We firstly conduct ablation experiments to prove the effectiveness of each proposed sub-modules on Voice Bank and DEMAND dataset~\cite{valentini2016investigating}. Specifically, the source speech comes from the VoiceBank corpus~\cite{veaux2013voice}, which contains 28 speakers for training and another 2 speakers for testing. Ten noise types with two artificially generated and eight real recordings from DEMAND~\cite{thiemann2013diverse} are used for training. Note that all data are downsampled from 48 K to 32K Hz before experimentation.
Totally, the fixed training %fixed training代表测试集是数据集里提前造好的 参考PERCEPTUAL LOSS BASED SPEECH DENOISING WITHAN ENSEMBLE OF AUDIO PATTERN RECOGNITION AND SELF-SUPERVISED MODELS
and validation set contains 11,572 utterances (10 h), and 872 utterances (30 min), respectively. %The test includes 824 samples, rated by 8 listeners each.
We also compare other SOTA models (including PercepNet~\cite{valin2020perceptually}) with \textit{S}-DCCRN on this dataset.

% 这块没说是咋训的呀？10h的集是提前造好的吗还是动态？信噪比？
%We first deliver ablation studies of the proposed model on Voice Bank and DEMAND dataset.
Then \textit{S}-DCCRN is further trained and evaluated with the Interspeech 2021 DNS challenge dataset to show its performance on more complicated and real acoustic scenarios.
%咋一会on Voice Bank and DEMAND dataset 一会又dns数据？
The source speech data comes from DNS-2021 full-band dataset, which contains 672 h speech data. The 180-hour DNS-2021 noise set, which includes 65,000 noise clips from 150 noise classes, is selected as the source noise data. The training set contains 605 h source speech data, while the validation set contains 67 h source speech data respectively. The training data are generated on-the-fly with 32K Hz sampling rate and segmented into 8 s chunks in one batch with SNR ranges from -5 to 20 dB. The total data `seen' by the model is more than 9000 h after 14 epochs of training.

%时长？
%具体数据细节直接跟在上一段说用了什么数据后面。你看一下我论文咋写的吧。
%For the noise set, the training set contains 40 different noise conditions with 10 types of noises (8 from the DEMAND database and 2 artificially generated)
%时长？
%at SNRs of 0 dB, 5 dB, 10 dB, and 15dB.
%这几个信噪比是啥意思？是你训练时候造的数据还是这个数据集本身的数据？如果是训练参数那就不是在这该说的 在下面生成数据时候说

%没看懂。。。

%we drop the utterances which duration is longer than 15 seconds because the IO of long utterances is slow.%啊？为啥？这跟你模型有善关系？Totally,
\vspace{-0.3cm}
\subsection{Training setup and baselines}
\vspace{-0.2cm}
For the proposed models, the window length and frame shift are 15ms and 5ms, respectively, resulting in a 20 ms at runtime of the model. The STFT length is 512.
For the models trained on Voice Bank and DEMAND dataset, all models are trained for 36 epochs with the following learning rate schedule: a constraint learning rate of 0.000025 is used for the first 10 epochs to warmup; then the learning rate is reset to 0.001. For the models trained on DNS-2021 dataset, the initial learning rate is 0.001 and will get halved if there is no loss decrease on the validation set. %~\cite{he2016deep}.
%for the rest epochs. For the models trained on DNS-2021 dataset, the initial learning rate is 0.001. When the loss of the validation set increases, the learning rate will be decayed by a ratio of 0.5.%valudation哪冒出来的?上面有没有说？看我论文咋说的
We also compare the proposed \textit{S}-DCCRN model and its ablation components on the Voice Bank and DEMAND dataset with other SOTA models. They are described as follows.

        \textbf{DCCRN}: The number of channels for the DCCRN is \{16,32,64,\\128,256,256\}, and the convolution kernel and step size are set to (5,2) and (2,1) respectively. Two LSTM layers are adopted and the number of nodes is 256. There is a 1024 $\times$ 256 fully connected layer after the LSTM. Each encoder/decoder module handles the current frame and one previous frame.

        \textbf{\textit{S}-DCCRN}: The number of channels for the sub-band DCCRN is \{32,64,64,64,128,128\}, and the convolution kernel and step size are set to (5,2) and (2,1) respectively. In addition, the channel number of the first layer of full-DCCRN is 64. One LSTM layer is adopted by sub-band DCCRN and full-band DCCRN respectively and the number of nodes is 256. There is a 256 $\times$ 256 fully connected layer after the LSTM. Each encoder/decoder module handles the current frame and one previous frame. The hidden channels of the complex feature encoder/decoder module are 32, and the depth of DenseBlock is 5. LayerNorm and PReLU are performed after each convolution, except for the last layer of the CFD module.

%小数据结果+可视化分析（曲线+不同语谱图） + 大数据结果
\vspace{-0.3cm}
\subsection{Experimental results and discussion}
As presented in Table~\ref{tab:vctkpesq}, ablation studies are conducted to evaluate the effectiveness of different model components of \textit{S}-DCCRN, including a) sub-band processing module (SP), b) sub-band and full-band processing module (SAF), c) \textit{S}-DCCRN without complex feature encoder/decoder module (CFE/CFD), d) \textit{S}-DCCRN without CFE/CFD and substitute learnable spectrum compression (LSC) with spectrum compression (SC), e) substitute LSC with SC. %we compare the contributions of the , , ,  and .
It should be noted that the SP only consists of a sub-band DCCRN, and thus the number of channels of encoder and decoder are \{2,64,64,128,128,256\}. The LSTM layers and units of SP are 2 and 256 respectively. We adopt CSIG, COVL, CBAK~\cite{hu2007evaluation}, STOI~\cite{taal2011algorithm} and PESQ~\cite{rix2001perceptual} as five evaluation metrics.
% 对比不同模型用abcd做序号 看我论文咋写的。而且你介绍sp这些的时候我感觉你这表达不对呀 sp是不是意思就是只一个sub-band dccrn？如果是那你就得说只包含了sub-band dccrn。后面如果是加上cfe cfd这些那就是说sub-band 和full-band加上了cfe cfd。不然读者还以为你这些模块是独立验证的呢。

    \begin{table}[!h]
    \centering
     \vspace{-0.6cm}
    \footnotesize
    \setlength\tabcolsep{2pt}% 调整列间距
\caption{Results of various models and ablation experiments on Voice Bank and DEMAND set.}
    \vspace{0.2cm}
    \begin{tabular}{lccccccc}
    \toprule
    Model   & \# Para.(M) & PESQ &CSIG & COVL & CBAK & STOI \\
    \midrule
    Noisy           & -  & 1.97  & 3.35 & 2.63    & 2.44 & 0.921  \\
    RNNoise           & 0.06  & 2.34  & 3.40 & 2.84    & 2.51 & 0.922  \\
    PercepNet           & 8  & 2.73  & - & -    & - &-  \\
    DCCRN           & 3.7  & 2.54  & 3.74 & 3.13    & 2.75 & 0.938  \\
    SP             & 2.76  & 2.63  & 3.86 & 3.23 & 3.03 &  0.935                                            \\
    SAF           & 2.73  & 2.71  & 3.94 & 3.31    & 3.08 & 0.937  \\
    ~ + SC           & 2.73  & 2.76  & 3.98 & 3.36   & 2.87 & 0.938  \\
    ~ + LSC           & 2.73  & 2.77  & 3.98 & 3.35    & 2.92 & 0.938  \\
    ~ + CFE/CFD          & 2.34  & 2.69  & 3.90 & 3.28    & \textbf{3.08} & 0.939  \\
    ~ ~ + SC         & 2.34  & 2.77  & 3.98 & 3.37    & 2.87 & 0.940  \\
    ~ ~ + LSC (\textit{S}-DCCRN)         & 2.34  & \textbf{2.84}  & \textbf{4.03} & \textbf{3.43} & 2.97  & \textbf{0.940}  \\
    \bottomrule
    \label{tab:vctkpesq}
    \vspace{-0.6cm}
    \end{tabular}
    \end{table}

    It can be seen from the results that the performance of the SAF module is obviously better than DCCRN, which obtains 0.17 PESQ improvement with a smaller model size.
    Compared with SP module, SAF module yielded 0.08 PESQ improvement. Because the low frequency and high frequency are modeled separately in the information of different frequency bands cannot be integrated. Adding CFE/CFD block yields lower PESQ compared to the SAF module alone. The low PESQ performance is caused by the low energy of high frequency, which is difficult for CFE/CFD block to model. This problem can be solved by adding a spectrum compression block. In addition, the results indicate that the proposed learnable spectrum compression is more effective than traditional spectrum compression~\cite{li2021importance} and it is especially beneficial for CFE/CFD block. It is worth to note that our approach achieves 0.11 PESQ improvement compared to Percepnet.
    %the highest PESQ score among RNNoise and Percepnet. %说一下领先了多少？
    \begin{figure}[b]
    \centering
    \vspace{-0.3cm}
    \includegraphics[width=.5\linewidth]{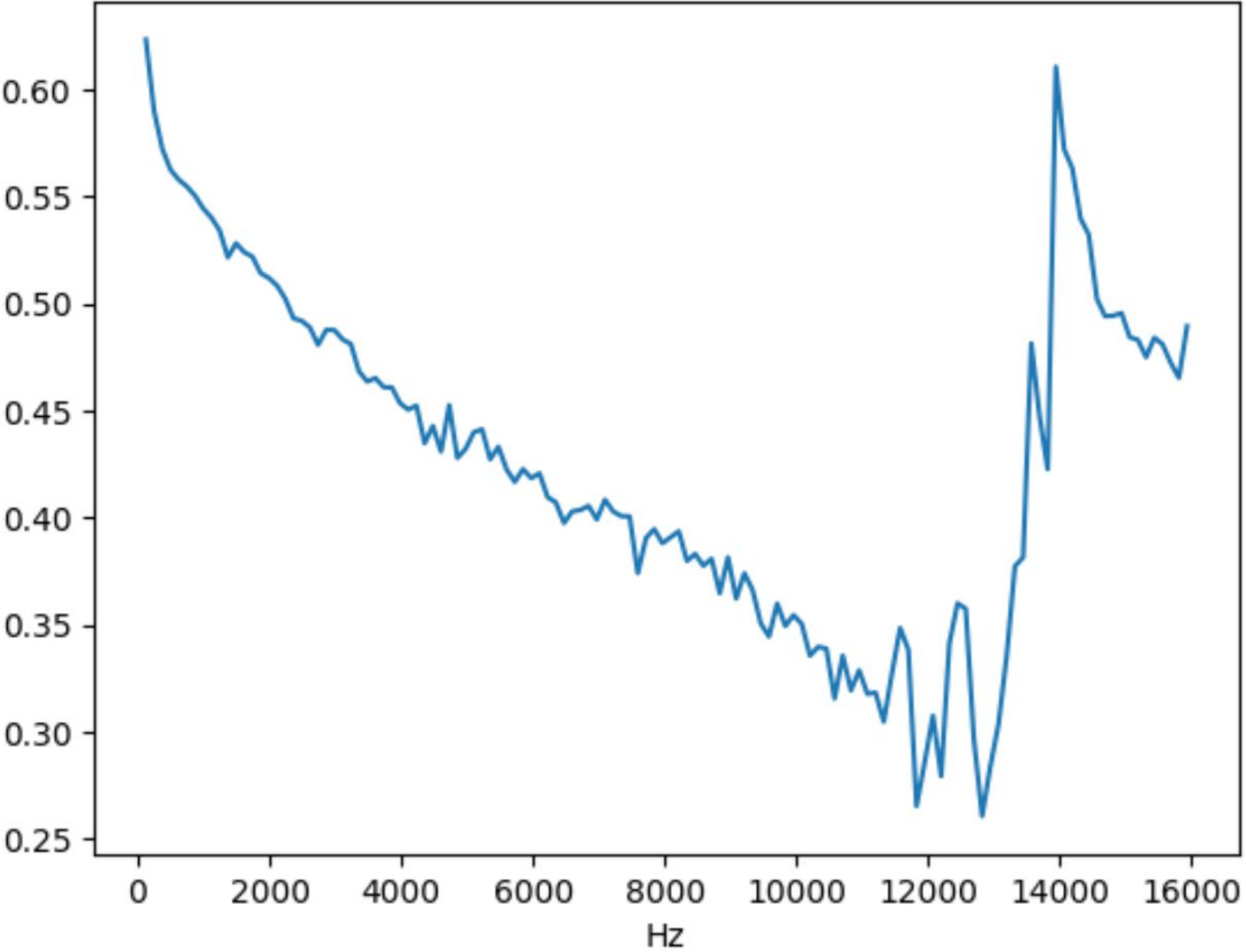}
     \vspace{-0.4cm}
    \caption{Compression ratio of different frequency automatically learned by the proposed learnable spectrum compression.}
    \label{fig:compressratio}
    \vspace{-0.4cm}
    \end{figure}

\begin{figure}[!h]
     \vspace{-0.2cm}
    \centering
    \includegraphics[width=1.0\linewidth]{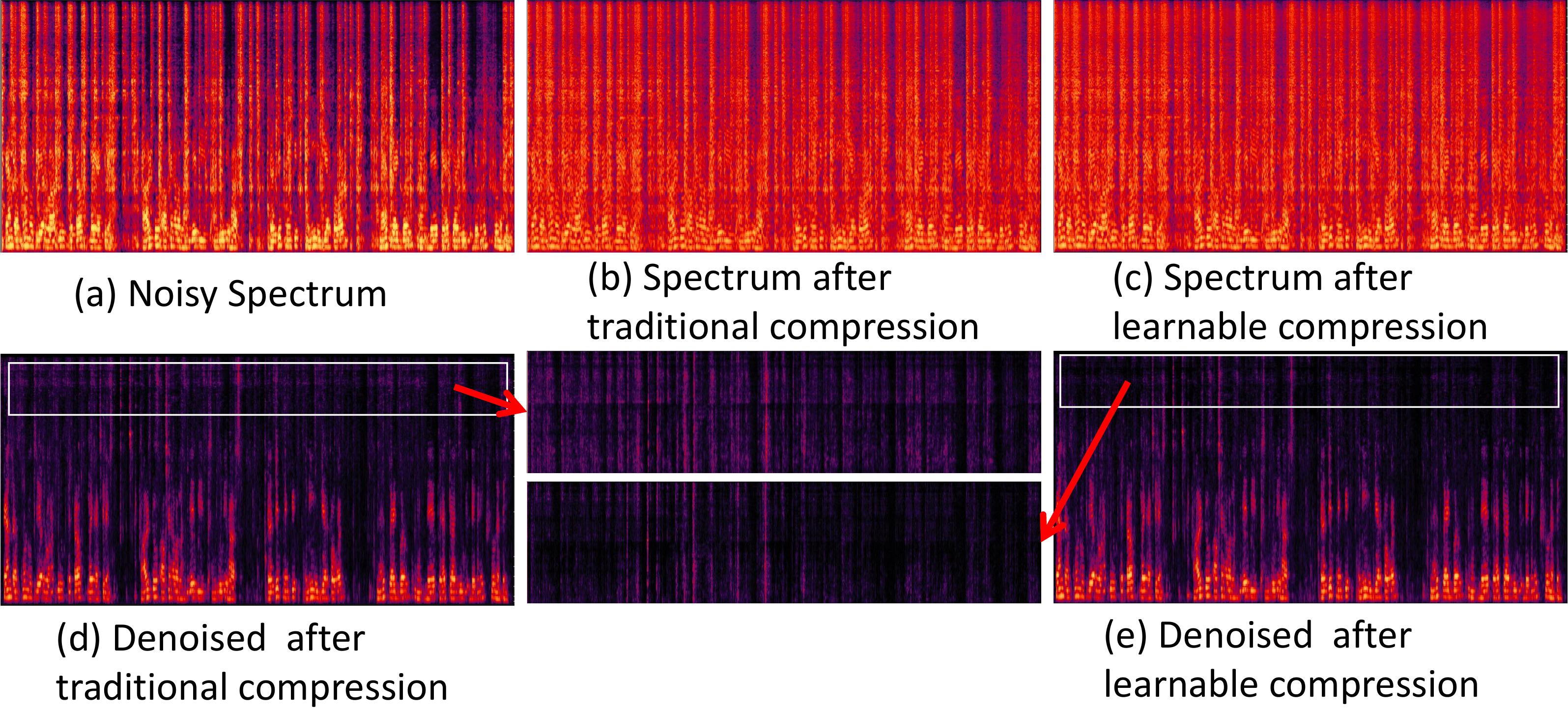}
    \vspace{-0.7cm}
    \caption{Comparison on the denoising result on a testing noisy clip for the cases with/without learnable spectrum compression.}
    \label{fig:speccompress}
    \vspace{-0.4cm}
    \end{figure}

    In Figure~\ref{fig:compressratio}, we show the learned compression ratio of different frequency bands on our denoising systems. We can observe that for the frequency bands lower than 13K Hz, the spectrum compression rate gradually decreases as the frequency increases. When frequency is higher than 13K Hz, the compression rate first increases and then stabilizes at about 0.5. We can consider that the low-frequency band has relatively larger energy and is unnecessary to be compressed to a great extent.
    %hard compressed
    In addition, the high-frequency bands demand a lower compression ratio to amplify the energy. Furthermore, the information of higher frequency bands is relatively few, which is not necessary to be deeply compressed. The spectrum after applying the traditional spectrum compression~\cite{li2021importance} and learnable spectrum compression are shown in Figure~\ref{fig:speccompress}. With the help of the learnable approach, the components of low-frequency band is not suppressed too much due to the high compression rate. Additionally, the speech and noise in the high-frequency band are more clear, leading to better noise reduction. %which is beneficial for \textit{S}-DCCRN to suppress the noise of high-frequency.
    %图和字都太小了 而且fig标题里也得说左中右分别是啥。

    %%%
    We further evaluate the models trained on the DNS-2021 dataset. Subjective tests are conducted by 10 listeners with aired hearing to evaluate the speech quality and intelligibility, in terms of 5-point mean opinion score (MOS)~\cite{recommendation2006vocabulary}, i.e, 1-bad, 2-poor, 3-fair, 4-good, 5-excellent on randomly selected 20 utterances. We also downsample the enhanced waveform to 16K Hz and conduct a DNSMOS evaluation~\footnote{DNSMOS only supports 16k Hz sampling rate}~\cite{reddy2021dnsmos}, which is used to simulate the human subjective evaluation. As shown in Table~\ref{tab:dnsmos}, with more training data, the MOS and DNSMOS for the SAF module are restored to the same level as DCCRN. When CFE/CFD and LSC modules are applied, the \textit{S}-DCCRN further improves the scores and achieves state-of-the-art performance.
    %%
    %%%DNSMOS测试集上模型效果对比
    \vspace{-0.6cm}
    \begin{table}[!h]
    \centering
    \footnotesize
    \setlength\tabcolsep{13pt}% 调整列间距
    \caption{MOS and DNSMOS results on DNS-2021 blind test set.}
    \vspace{0.3cm}
    \begin{threeparttable}
    \begin{tabular}{lccc}
    \toprule
    Model   & MOS &DNSMOS$^*$ \\
    \midrule
    Noisy     & 1.66 & 2.94 \\
    RNNoise & 2.32 & 3.07 \\
    DCCRN  & 3.30 & 3.31 \\
    SAF & 3.20 & 3.33 \\
    \textit{S}-DCCRN & \textbf{3.62} & \textbf{3.43} \\
    \bottomrule
    \label{tab:dnsmos}
    \end{tabular}
    %\begin{tablenotes}
     \begin{tablenotes}
            %\footnotesize
            \vspace{-0.3cm}
            \item *: Calculated on downsampled speech (16K Hz)  %此处加入注释*信息 %此处加入注释**信息
          \end{tablenotes}
    \end{threeparttable}

    %\end{tablenotes}
    %\vspace{-0.7cm}
    \end{table}
    \vspace{-0.7cm}
\section{Conclusions}
\label{sec:pagestyle}
\vspace{-0.3cm}
In this paper, we propose a novel super wide-band STFT domain denoising network runing on 32K signal. The \textit{S}-DCCRN is equipped with SAF module via a cascaded sub-band and full-band processing module, aiming at benefiting from both local and global frequency information processing. Importantly, a complex feature encoder/decoder is adopted to refine the information of different frequency bands. Finally, a learnable spectrum compression method is employed to adjust the energy of different frequency bands. The proposed \textit{S}-DCCRN model obtains superior performance with 3.62 MOS score on the blind test set of Interspeech 2021 DNS challenge. Experiments have shown the effectiveness of these methods~\footnote{Demo page is available at https://imybo.github.io/S-DCCRN/}.

% \section{REFERENCES}
% \label{sec:refs}

% List and number all bibliographical references at the end of the
% paper. The references can be numbered in alphabetic order or in
% order of appearance in the document. When referring to them in
% the text, type the corresponding reference number in square
% brackets as shown at the end of this sentence \cite{C2}. An
% additional final page (the fifth page, in most cases) is
% allowed, but must contain only references to the prior
% literature.

\bibliographystyle{IEEE}
\bibliography{reference}

\end{document}